\documentclass[12pt]{article}

\usepackage{amsmath}   % From the American Mathematical Society
\usepackage{amsthm}    % A package that provides \newtheorem command
\usepackage{amssymb}
\usepackage{graphics}  % Simple including of graphics, photos, etc.
\usepackage{graphicx}
\usepackage{color}
\usepackage{multirow}

\DeclareMathOperator*{\argmin}{argmin}

\begin{document}

%%% Title section
\begin{center}
{\Large\bf The spatial sign covariance matrix and its application for robust correlation estimation}
\end{center}
\begin{center}
{\sc A. D\"urre\footnote{corresponding author, e-mail: {\tt alexander.duerre@udo.edu}}, R. Fried}\\
{\it Fakult\"at Statistik, Technische Universit\"at Dortmund\\
44221 Dortmund, Germany\\}
{\sc D. Vogel}\\
{\it Institute for Complex Systems and Mathematical Biology, University of Aberdeen\\
Aberdeen AB24 3UE, United Kingdom\\}
\end{center}

%%% abstract
\begin{abstract}
We summarize properties of the spatial sign covariance matrix and especially look at the relationship between its eigenvalues and those of the shape matrix of an elliptical distribution. The explicit relationship known in the bivariate case was used to construct the spatial sign correlation coefficient, which is a non-parametric and robust estimator for the correlation coefficient within the elliptical model. We consider a multivariate generalization, which we call the multivariate spatial sign correlation matrix.
\end{abstract}

%%%
\section{Introduction}
Let $\mathbf{X}_1,\ldots,\mathbf{X}_n$ denote a sample of independent $p$ dimensional random variables from a distribution $F$ and $s: \mathbb{R}^p\rightarrow \mathbb{R}^p$ with $s(\mathbf{x})=\mathbf{x}/|\mathbf{x}|$ for $\mathbf{x}\neq 0$ and $s(0)=0$ the spatial sign, then
\begin{align*}
S_n(\mathbf{t}_n,\mathbf{X}_1,\ldots,\mathbf{X}_n)=\frac{1}{n}\sum_{i=1}^n s(\mathbf{X}_i-\mathbf{t}_n)s(\mathbf{X}_i-\mathbf{t}_n)^T
\end{align*}
denotes the empirical spatial sign covariance matrix (SSCM) with location $\mathbf{t}_n$.
The canonical choice for the location estimator $\mathbf{t}_n$ is the spatial median 
\begin{align*}
\boldsymbol{\mu}_n=\argmin_{\boldsymbol{\mu} \in \mathbb{R}^p} \sum_{i=1}^n ||\mathbf{X}_i-\boldsymbol{\mu}||.
\end{align*}
Beside its nice robustness properties like an asymptotic breakdown-point of 1/2, it has (under regularity conditions, see \cite{Kemperman:spatialmedian}) the advantageous feature that it centres the spatial signs, i.e.,
\begin{align*}
\frac{1}{n}\sum_{i=1}^n s(\mathbf{X}_i-\boldsymbol{\mu}_n)=0,
\end{align*}
so that $S_n(\boldsymbol{\mu}_n,\mathbf{X}_1,\ldots,\mathbf{X}_n)$ is indeed the empirical covariance matrix of the spatial signs of the data.
If $\mathbf{t}_n$ is (strongly) consistent for a location $\mathbf{t}\in \mathbb{R}$, it was shown in \cite{duerre:vogel:tyler:sscm} that under mild conditions on $F$ the empirical SSCM is a (strongly) consistent estimator for its population counterpart
\begin{align*}
S(\mathbf{X})=\mathbb{E}(s(\mathbf{X}-\mathbf{t})s(\mathbf{X}-\mathbf{t})^T).
\end{align*}
There are some nice results if $F$ is within the class of continuous elliptical distributions, which means that $F$ possesses a density of the form
\begin{align*}
f(\mathbf{x})=\mbox{det}(V)^{-\frac{1}{2}}g((\mathbf{x}-\boldsymbol{\mu})V^{-1}(\mathbf{x}-\boldsymbol{\mu}))
\end{align*} 
for a location $\boldsymbol{\mu}\in \mathbb{R}^p$, a symmetric and positive definite shape matrix $V\in \mathbb{R}^{p\times p}$ and a function $g:\mathbb{R}\rightarrow \mathbb{R}$, which is often called the elliptical generator. 
Prominent members of the elliptical family are the multivariate normal distribution and elliptical $t$-distributions (e.g.\ \cite{Bilodeau1999}, p.~208). 
If second moments exists, then $\boldsymbol\mu$ is the expectation of $\mathbf{X} \sim F$, and $V$ a multiple of the covariance matrix. The shape matrix $V$ is unique only up to a multiplicative constant. 
In the following, we consider the trace-normalized shape matrix $V_0=V/\mbox{tr}(V),$ which is convenient since $S(\mathbf{X})$ also has trace 1. If $F$ is elliptical, then $S(\mathbf{X})$ and $V$ share the same eigenvectors and the respective eigenvalues have the same ordering. 
For this reason, the SSCM has been proposed for robust principal component analysis (e.g.\  \cite{locantore:robpca, marden:robpca}). In the present article, we study the eigenvalues of the SSCM.

\section{Eigenvalues of the SSCM}
Let $\lambda_1\geq \ldots \geq \lambda_p\geq 0$ denote the eigenvalues of $V_0$ and $\delta_1\geq \ldots \geq \delta_p\geq 0$ those of $S(\mathbf{X})$.
Explicit formulae that relate the $\delta_i$ to the $\lambda_i$ are only known for $p=2$ (see \cite{vogel:parcor, croux:ksscm}), namely
\begin{align}\label{eveq}
\delta_i=\frac{\sqrt{\lambda}_i}{\sqrt{\lambda}_1+\sqrt{\lambda}_2},~~~i=1,2.
\end{align}
Assuming $\lambda_2>0$, we have 
$\delta_1/\delta_2=\sqrt{\lambda_1/\lambda_2}\leq \lambda_1/\lambda_2,$
thus the eigenvalues of the SSCM are closer together than those of the corresponding shape matrix. It is shown in \cite{durre:evsscm} that this holds true for arbitrary $p>2$, so 
\begin{align}\label{evrel}
\lambda_i/\lambda_j\geq \delta_i/\delta_j~~~\mbox{for } 1\leq i < j \leq p
\end{align}
as long as $\lambda_j>0.$ 
There is no explicit map between the eigenvalues known for $p>2$. D\"urre et al.\ \cite{durre:evsscm} give a representation of $\delta_i$ as one-dimensional integral, which permits fast and accurate numerical evaluations for arbitrary $p$, 
\begin{align}\label{evequation}
\delta_i=\frac{\lambda_i}{2}\int_0^\infty \frac{1}{(1+\lambda_ix)\prod_{j=1}^p(1+\lambda_jx)^\frac{1}{2}}dx, ~~~i=1,\ldots,p.
\end{align}
We use this formula (implemented in R \cite{R} in the package sscor \cite{sscor}) to get an impression how the eigenvalues of $S(\mathbf{X})$ look like in comparison to those of $V_0$. We first look at of equidistantly spaced eigenvalues
\begin{align*}\lambda_i=\frac{2i}{p(p+1)},~~i=1,\ldots,p, 
\end{align*} for different $p=3,~11,~101$.
\begin{figure}
\begin{center}
\includegraphics[width=0.99 \textwidth ]{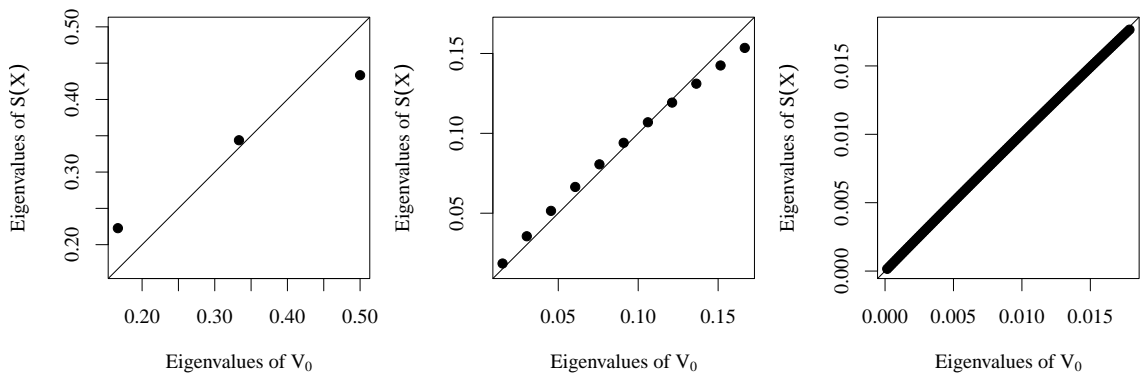}
\caption{Eigenvalues of the SSCM wrt the corresponding eigenvalues of the shape matrix in the equidistant setting $p=3$ (left), $p=11$ (centre) and $p=101$ (right).}
\label{eval}
\end{center}
\end{figure}
The magnitude of the eigenvalues necessarily decreases as $p$ increases, since $\sum_{i=1}^p \lambda_i=\sum_{i=1}^p \delta_i=1$ per definition of $V_0$ and $S(\mathbf{X})$. As one can see in Figure \ref{eval}, the eigenvalues of $S(\mathbf{X})$ and $V_0$ approach each other for increasing $p$. In fact the maximal absolute difference for $p=101$ is roughly $2\cdot 10^{-4}$. 
In the second scenario, we take $p-1$ equidistantly spaced eigenvalues and one eigenvalue 5 times larger than the rest, i.e.,
\begin{align*}
\lambda_i=\begin{cases}\frac{i}{p((p+1)/{2}+5)-5} &i=1,\ldots,p-1, \\[1.0ex]
\frac{5(p-1)}{p((p+1)/{2}+5)-5}&i=p.
\end{cases}
\end{align*}
This models the case where the dependence is mainly driven by one principle component. 
\begin{figure}
\begin{center}
\includegraphics[width=0.99 \textwidth]{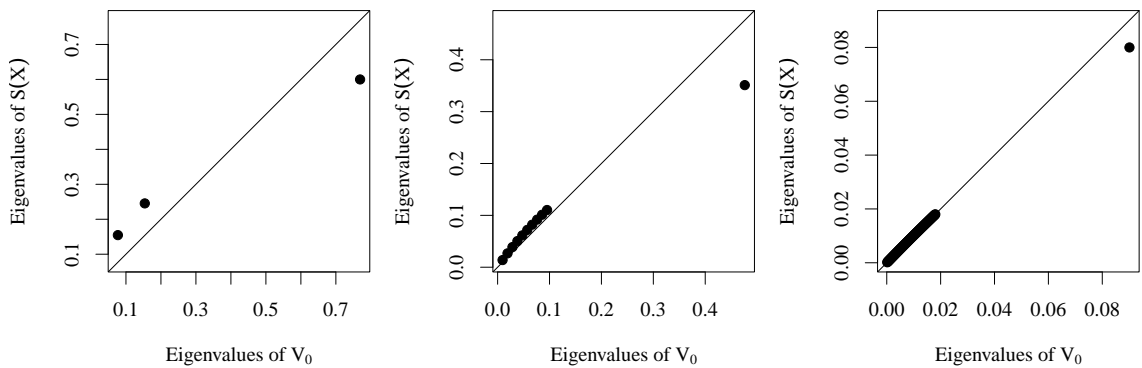}
\caption{Eigenvalues of the SSCM wrt the corresponding eigenvalues of shape matrix in the setting of one large eigenvalue for $p=3$ (left), $p=11$ (centre) and $p=101$ (right).}
\label{eval2}
\end{center}
\end{figure}
As one can see in Figure \ref{eval2}, the distance between the two largest eigenvalues is smaller for $S(\mathbf{X})$ than for $V_0$. This is not surprising in light of (\ref{evrel}). 
Thus in general, the eigenvalues of the SSCM are less separated than those of $V_0$, which is one reason why the use of the SSCM for robust principal component analysis has been questioned (e.g.\ \cite{bali, magyar}). However, the differences appear to be generally small in higher dimensions.

\section{Estimation of the correlation matrix}

Equation (\ref{eveq}) can be used to derive an estimator for the correlation coefficient based on the empirical SSCM: the spatial sign correlation coefficient $\rho_n$ (\cite{durre:sscor}). Under mild regularity assumptions this estimator is consistent under elliptical distributions and asymptotically normal with variance
\begin{align}\label{asi1}
\mbox{ASV}(\rho_n)=(1-\rho^2)^2+\frac{1}{2}(a+a^{-1})(1-\rho^2)^{3/2},
\end{align} 
where $a=\sqrt{v_{11}/v_{22}}$ is the ratio of the marginal scales and $\rho=v_{12}/\sqrt{v_{11}v_{22}}$ is the generalized correlation coefficient, which coincides with the usual moment correlation coefficient if second moments exists. Equation (\ref{asi1}) indicates that the variance of $\rho_n$ is minimal for $a=1$, but can get arbitrarily large if $a$ tends to infinity or 0.

Therefore a two-step procedure has been proposed, the \emph{two-stage spatial sign correlation} $\rho_{\sigma,n}$, which first normalizes the data by a robust scale estimator, e.g., the median absolute deviation (mad), and then computes the spatial sign correlation of the transformed data. Under mild conditions (see \cite{durre:sscorts}), this two-step procedure yields an asymptotic variance of
\begin{align}\label{asi2}
\mbox{ASV}(\rho_{\sigma,n})=(1-\rho^2)^2+(1-\rho^2)^{3/2},
\end{align}
which equals that of $\rho_n$ for the favourable case of $a=1$. 
Since (\ref{asi2}) only depends on the parameter $\rho$, the two-stage spatial sign correlation coefficient is very suitable to construct robust and non-parametric confidence intervals for the correlation coefficient under ellipticity. It turns out that these intervals are quite accurate even for rather small sample sizes of $n=10$ and in fact more accurate then those based on the sample moment correlation coefficient \cite{durre:sscorts}.

One can construct an estimator of the correlation matrix $R$ by filling the off-diagonal positions of the matrix estimate with the bivariate spatial sign correlation coefficients of all pairs of variables. This was proposed in \cite{durre:sscor}. 
Equation (\ref{evequation}) allows an alternative approach: First standardize the data by a robust scale estimator and compute the SSCM of the transformed data. Then apply a singular value decomposition
\begin{align*}
S_n(\mathbf{t}_n,\mathbf{X}_1,\ldots,\mathbf{X}_n)=\hat{U}\hat{\Delta}\hat{U}^T,
\end{align*}
where $\hat{\Delta}$ contains the ordered eigenvalues $\hat{\delta}_1\geq \ldots \geq \hat{\delta}_p.$ One obtains estimates $\hat\lambda_1,\ldots, \hat\lambda_p$ by inverting (\ref{evequation}). Although theoretical results are yet to be established, we found in our simulations that the following fix point algorithm
\begin{align*}
	\hat{\lambda}_i^{(0)} 	& =  \delta_i,       & & \ i=1,\ldots,p,  \\ 
	\tilde{\lambda}_i^{(k+1)} & =  2\hat{\delta}_i 
	\left( 
			\int_0^\infty \frac{1}{(1+\hat{\lambda}_i^{(k)}x)\prod_{j=1}^p(1+\hat{\lambda}^{(k)}_j x)^\frac{1}{2}}dx, 
	\right)^{-1},  & & \ i=1,\ldots,p, \ k=1,2,\ldots \\
	\hat{\lambda}_i^{(k+1)} & =\tilde{\lambda}_i^{(k+1)}\left(\sum_{j=1}^p\tilde{\lambda_j}^{(k+1)}\right)^{-1},  & & \ i=1,\ldots,p, \ k=1,2,\ldots 
\end{align*}
works reliably and converges fast. Let $\hat{\Lambda}$ denote the diagonal matrix containing $\hat{\lambda}_1,\ldots,\hat{\lambda}_p,$ then 
%\begin{align*}
$\hat{V}=\hat{U}\hat{\Lambda}\hat{U}^T$ 
%\end{align*}
is a suitable estimator for for the shape of the standardized data and 
%\begin{align*}
$\hat{R}$ with $\hat{r}_{ij}=\hat{v}_{ij}/\sqrt{\hat{v}_{ii}\hat{v}_{jj}}$
%\end{align*}
an estimator for the correlation matrix, which we call the \emph{multivariate spatial sign correlation matrix}. Contrary to the pairwise approach, the multivariate spatial sign correlation matrix is positive semi-definite by construction.

Theoretical properties of the new estimator are not straightforward to establish.
By a small simulation study we want to get an impression of its efficiency. 
We compare the variances of the moment correlation, the pairwise as well as the multivariate spatial sign correlation under several elliptical distributions: normal, Laplace and $t$ distributions with 5 and 10 degrees of freedom. The latter three generate heavier tails than the normal distribution. 
The Laplace distribution is obtained by the elliptical generator $g(x) = c_p \exp(-\sqrt{|x|}/2)$, where $c_p$ is the appropriate integration constant depending on $p$ (e.g.\ \cite{Bilodeau1999}, p.~209).

We take the identity matrix as shape matrix and compare the variances of an off-diagonal element of the matrix estimates for different dimensions $p=2,~3,~5,~10,~50$ and sample sizes $n=100,~1000$. We use the R packages mvtnorm \cite{mvtnorm} and MNM \cite{MNM} for the data generation. The results based on 10000 runs are summarized in Table \ref{effit}.
\begin{table}\begin{center}
 \begin{tabular}{ll|ccccc|ccccc}
 \multicolumn{2}{l|}{\hfill$n$}&\multicolumn{5}{c|}{$100$}&\multicolumn{5}{c}{$1000$}\\
 \multicolumn{2}{l|}{\hfill$p$}&2&3&5&10&50&2&3&5&10&50\\ \hline\hline
 \multirow{3}{*}{$N$}&\mbox{cor}&1.0&1.0&1.0&1.0&1.0&1.0&1.0&1.0&1.0&1.0\\
 &\mbox{sscor pairwise}                 &1.9&1.9&1.9&1.9&1.9&2.0&2.0&2.0&2.0&2.0\\
 &\mbox{sscor multivariate}                 &1.9&1.6&1.4&1.2&1.0&2.0&1.7&1.4&1.2&1.0\\ \hline
 \multirow{3}{*}{$t_{10}$}&\mbox{cor}&1.3&1.3&1.3&1.3&1.3&1.3&1.3&1.3&1.4&1.3\\
  &\mbox{sscor pairwise}                     &2.0&1.9&1.9&2.0&1.9&2.0&2.0&2.0&2.0&2.0\\
  &\mbox{sscor multivariate}                     &2.0&1.7&1.3&1.2&1.0&2.0&1.7&1.4&1.2&1.0\\ \hline
 \multirow{3}{*}{$t_5$}&\mbox{cor} &2.0&2.1&2.1&2.1&2.1&2.6&2.6&2.6&2.6&2.6\\
  &\mbox{sscor pairwise}                   &2.0&2.0&1.9&2.0&1.9&2.1&2.0&2.0&2.0&2.0\\
  &\mbox{sscor multivariate}                   &2.0&1.7&1.4&1.2&1.1&2.1&1.7&1.4&1.2&1.0\\ \hline
 \multirow{3}{*}{$L$}&\mbox{cor}   &1.6&1.5&1.3&1.2&1.1&1.6&1.5&1.3&1.2&1.1\\
 &\mbox{sscor pairwise}                    &1.9&1.9&1.9&2.0&2.0&2.0&2.0&2.0&2.0&2.0\\
 &\mbox{sscor multivariate}                    &1.9&1.6&1.4&1.2&1.1&2.0&1.7&1.4&1.2&1.1
\end{tabular}
\caption{Simulated variances (multiplied by $\sqrt{n}$) of one off-diagonal element of the correlation matrix estimate based on the moment correlation (cor), the pairwise spatial sign correlation (sscor pairwise) and the multivariate spatial sign correlation matrix (sscor multivariate) for spherical normal ($N$), $t_5$, $t_{10}$, and Laplace ($L$) distribution, several dimensions $p$ and sample sizes $n = 100, 1000$.\label{effit}}
 \end{center}
\end{table}

Except for the moment correlation at the $t_5$ distribution, the results for $n=100$ and $n=1000$ are very similar.
Note that the variance of the moment correlation decreases at the Laplace distribution as the dimension $p$ increases, but not so for the other distributions considered. The lower dimensional marginals of the Laplace distribution are, contrary to the normal and the $t$-distributions, not Laplace distributed (see \cite{Kano1994}), and the kurtosis of the one-dimensional marginals of the Laplace distribution in fact decreases as $p$ increases.

Equation (\ref{asi2}) yields an asymptotic variance of 2 for the pairwise spatial sign correlation matrix elements regardless of the specific elliptical generator, which can also be observed in the simulation results. 
The moment correlation is twice as efficient under normality, but has a higher variance at heavy tailed distributions. 
For uncorrelated $t_5$ distributed random variables, the spatial sign correlation outperforms the moment correlation. 
Looking at the multivariate spatial sign correlation, we see a strong increase of efficiency for larger $p$. For $p=50$ the variance is comparable to that of the moment correlation. 
Since the asymptotic variance of the SSCM does not depend on the elliptical generator, this is expected to also hold for the multivariate spatial sign correlation, and we find this confirmed by the simulations. The multivariate spatial sign correlation is more efficient than the moment correlation even under slightly heavier tails for moderately large $p$.

An increase of efficiency for larger $p$ is not uncommon for robust scatter estimators. It can be observed amongst others for $M$-estimators, the Tyler shape matrix, the MCD, and $S$-estimators (e.g. \cite{croux:inf, Taskinen}). 
All of these are affine equivariant estimators, requiring $n > p$. This is not necessary for the spatial sign correlation matrix. One may expect that the efficiency gain for large $p$ is at the expense of robustness, in particular a larger maximum bias curve. 
Further research will be necessary to thoroughly explore the robustness properties and efficiency of the multivariate spatial sign correlation estimator.

%%%

%Here is the text of the section. 

%The command \verb|\providecommand| should be used instead of
%the command \verb|\newcommand|. 

%The command \verb|\provideenvironment| 
%should be used instead of the command \verb|\newenvironmant|. 

% An example of a mathematical environments
%\begin{theorem}
%The sample of a theorem.
%\end{theorem}
%\begin{lemma}
%The sample of a lemma.
%\end{lemma}
%\begin{corollary}
%The sample of a corollary.
%\end{corollary}

% An example of a floating figure using the graphics package.
%\begin{figure}
%\begin{center}
%\includegraphics{myfigure.eps}
%\caption{Simulation Results}
%\label{Author:fig_sim}
%\end{center}
%\end{figure}

% An example of a floating table.
%\begin{table}
%\begin{center}
%\caption{An Example of a Table}
%\label{table_example}
%\begin{tabular}{|c|c|}
%\hline
%One & Two\\
%\hline
%Three & Four\\
%\hline
%\end{tabular}
%\end{center}
%\end{table}

\end{document}